# Ferroelectric Polarization Rotation in Order-Disorder-Type LiNbO$_3$ Thin Films

*Tae Sup Yoo[1,‡], Sang A Lee[1,‡], Changjae Roh[2], Seunghun Kang[3], Daehee Seol[3], Xinwei Guan[4], Jong-Seong Bae[5], Jiwoong Kim[6], Young-Min Kim[7,8], Hu Young Jeong[9], Seunggyo Jeong[1], Ahmed Yousef Mohamed[10], Deok-Yong Cho[10], Ji Young Jo[11], Sungkyun Park[6], Tom Wu[4,12], Yunseok Kim[3], Jongseok Lee[2], and Woo Seok Choi[1,\*]*

[1]Department of Physics, Sungkyunkwan University, Suwon 16419, Korea

[2]Department of Physics and Photon Science, Gwangju Institute of Science and Technology (GIST), Gwangju 61005, Korea

[3]School of Advanced Materials Science and Engineering, Sungkyunkwan University, Suwon 16419, Korea

[4]Materials Science and Engineering, King Abdullah University of Science and Technology, Thuwal 23955-6900, Saudi Arabia

[5]Busan Center, Korea Basic Science Institute, Busan 46742, Korea

[6]Department of Physics, Pusan National University, Busan 46241, Korea

[7]Department of Energy Science, Sungkyunkwan University, Suwon 16419, Korea

[8]Center for Integrated Nanostructure Physics, Institute for Basic Science (IBS), Suwon 16419, Korea

[9]UNIST Central Research Facilities, Ulsan National Institute of Science and Technology, Ulsan

44919, Korea

[10]IPIT & Department of Physics, Chonbuk National University, Jeonju 54896, Korea

[11]School of Materials Science and Engineering, Gwangju Institute of Science and Technology (GIST), Gwangju 61005, Korea

[12]School of Materials Science and Engineering, University of New South Wales, Sydney, NSW 2052, Australia

KEYWORDS.

ferroelectric polarization rotation, $LiNbO_3$ thin films, vacancy engineering, spatial inversion symmetry breaking, second harmonic generation.

ABSTRACT.

The direction of ferroelectric polarization is prescribed by the symmetry of the crystal structure. Therefore, rotation of the polarization direction is largely limited, despite the opportunity it offers in understanding important dielectric phenomena such as piezoelectric response near the morphotropic phase boundaries and practical applications such as ferroelectric memory. In this study, we report the observation of continuous rotation of ferroelectric polarization in order-disorder type $LiNbO_3$ thin films. The spontaneous polarization could be tilted from an out-of-plane to an in-plane direction in the thin film by controlling the Li vacancy concentration within the hexagonal lattice framework. Partial inclusion of monoclinic-like phase is attributed to the breaking of macroscopic inversion symmetry along different directions and the emergence of ferroelectric polarization along the in-plane direction.

**INTRODUCTION**

The ability to rotate the direction of spontaneous polarization in ferroelectrics provides an essential insight in understanding and utilizing the piezoelectric and ferroelectric phenomena. The rotation of the ferroelectric polarization is closely coupled with the ferroelectric instability and piezoelectric elongation,[1-2] which are imperative in improving the performance of ferroelectric sensors, energy-harvesting devices, and ferroelectric memories. As a prominent example, polarization rotation should necessarily occur at a morphotropic phase boundary (MPB), i.e., a border between two phases with different polarization directions, which would show an enhanced piezoelectric response.[1-6] Enhanced piezoelectricity was also observed in ferroelectric $BiFeO_3$ thin films doped with Co, where the polarization direction rotates concurrently with tetragonal/monoclinic structural phase transition.[7] In $PbZr_{1-x}Ti_xO_3$ thin films, epitaxial strain induced the rotation of the electric dipole resulting in structural twinning.[8-9] In $PbTiO_3/CaTiO_3$ superlattices, a large dielectric constant resulted at a specific tilt angle of the polarization direction, as the $CaTiO_3$ volume fraction was varied.[10] More recently, continuous rotation of the polarization was observed in ferroelectric superlattices which resulted in a polar vortex structure with converging dipoles.[11-12]

Despite the growing interest, the rotation of the polarization direction has been demonstrated only in conventional displacive-type ferroelectrics until today. In a displacive-type ferroelectric, the lowest energy coordinate position can shift continuously with structural modification, resulting in a systematic rotation of electric dipoles. In the order-disorder type ferroelectric, on the other hand, the lowest energy coordinate is relatively fixed even above the ferroelectric transition temperature ($T_C$), but the randomly pointing dipoles align below $T_C$ giving rise to the ferroelectricity. Thus, ferroelectric polarization rotation in an order-disorder type ferroelectric will manifest different underlying mechanisms from that of the displacive-type ferroelectric, and a close observation of such phenomenon is expected to complement our understanding of ferroelectricity.

LiNbO$_3$ (LNO) is known as a classic order-disorder type ferroelectric.[13] It shows large remnant polarization of ~70 $\mu$C/cm$^2$, piezoelectric response of ~6 pm/V in bulk, and features attractive optical properties such as strong birefringence.[13-17] Photovoltaic effect related to ferroelectricity has also been discussed in LNO, despite its wide band gap of ~4 eV.[15, 18] The LNO lattice structure belongs to $R3c$ space group at room temperature and consists of Li–O trigon and Nb–O octahedron along the $c$-axis of the hexagonal unit cell. Li ions play a dominant role in the ferroelectric behavior according to a first-principles calculation study.[19] The displacement of Li ions occurs either upward or downward in relation to the triangular oxygen layer, resulting in spontaneous polarization along only the $c$-axis direction. Nb ions also move within the octahedra, making an additional contribution to the ferroelectricity.

Bulk LNO crystals have been extensively studied and widely adopted in practical applications. However, growth and characterization of crystalline LNO thin film have been largely limited. Most of the limited number of studies on LNO thin films have adopted sapphire as substrates, despite a large lattice mismatch (−8.2%, compressive).[20-22] While it maintains the same in-plane hexagonal symmetry, the produced films exhibit poor crystallinity. LNO films have also been deposited on GaAs substrates, or using buffer layers such as ZnO and SiO$_2$ on Si substrates.[23-25] However, these studies did not present ideal crystalline quality or good ferroelectric behavior expected from LNO bulk crystals. Realization of LNO thin films with good crystallinity would certainly widen the possibilities for manipulating the ferroelectric properties and help us assess the fundamental opto-electronic behavior in LNO.

In this study, we aim to demonstrate rotation of the spontaneous polarization direction in the order-disorder type ferroelectric LNO thin film. The strong coupling between chemical potential and lattice structure is attributed to the rotation of the polarization. In particular, the unexpected polarization rotation from an out-of-plane direction to an in-plane direction of the LNO thin film could be understood in terms of inclusion of anti-site Nb defects and resultant local monoclinic $LiNb_3O_8$-like structure in possession of off-centered ions along the in-plane direction. As a result, a LNO thin film with fully in-plane ferroelectric polarization could be realized.

**EXPERIMENTAL DETAILS**

**Thin Film Growth.** We fabricated LNO (001) thin films by pulsed laser deposition (PLD) under various oxygen partial pressures ($P(O_2)$) of 5–50 mTorr at 600ºC. The thin films were deposited on Ti-terminated atomically flat $SrTiO_3$ (STO) (111), and Nb (0.5wt%)-doped STO (Nb:STO) (111) substrates (when a metallic bottom electrode was necessary).[26] A laser fluence of 1.5 J/cm$^2$ was used. The thickness of all thin films was approximately 75 nm as measured using scanning transmission electron microscopy (STEM).

**Lattice structure characterization.** X-ray diffraction (XRD) $\theta$-$2\theta$ and pole figure measurements were performed using high resolution x-ray diffractometer (Rigaku SmartLab and PANalytical X'Pert). For pole figure measurements, $2\theta$ positions were aligned to the Bragg reflection conditions of $SrTiO_3$ (110) and monoclinic $LiNb_3O_8$ ($\bar{2}12$) at $2\theta$ = 32.396º and 32.200º, respectively. Atomic-scale LNO structures were analyzed on a spherical aberration-corrected scanning transmission electron microscope (STEM, JEOL ARM200CF) operating at 200 kV. To resolve light elements such as oxygen and lithium in the structure, annular bright field (ABF) STEM imaging mode was employed. The incident electron probe angle was 24 mrad and the ABF signals were collected over a detector angle window of 7.5–17 mrad. TEM cross-sectional samples were prepared with the use of dual-beam focused ion beam system (FIB, FEI Helios Nano Lab 450) and consecutive low-energy Ar ion milling at 700 V (Fischione Model 1040 NanoMill) was carried out for 15 min to remove damaged surface layers caused by heavy Ga ion beam milling in the FIB system.

**X-ray spectroscopy.** Qualitative changes in the stoichiometry were evaluated by x-ray photoelectron spectroscopy (XPS, Thermo Scientific) and x-ray absorption spectroscopy (XAS) at room temperature. XPS was carried out using a monochromatic Al $K\alpha$ source (1486.6 eV) with step size of 0.05 eV, pass energy of 50.0 eV, and spot size of 400 μm. Nb $L_3$-edge XAS (~2370 eV) was performed in the 16A1 beamline at the Taiwan Light Source (TLS). The fluorescence yield was collected using a Lytle detector. The angle between the sample plane and incident x-ray was set to 60°. The photon energy was calibrated by the edge energy of an Nb metal foil.

**Second harmonic generation.** A second harmonic generation (SHG) experiment was performed at room temperature in order to confirm the structural inversion symmetry breaking. Fundamental light source was a 100-fs-long pulsed laser (Coherent Vitara-T) with a center wavelength at 800 nm and repetition rate of 80 MHz. Fundamental wave with average power of 100 mW was incident on the sample surface within a diameter of about 100 μm, and the SHG light intensity was detected using a photomultiplier tube (PMT) operating in specular reflection geometry with incidence angle of 45°. Fundamental light polarization was established using a half wave plate (HWP), and the polarization state of SHG light was analyzed by an additional polarizer placed just after the sample.

**Piezoresponse force microscopy.** Piezoresponse force microscopy (PFM) was conducted at room temperature for exploring anisotropic distribution of polarizations. Here, AC modulation bias of 1.0 V at 17 kHz was applied to a conductive probe (BudgetSensors Multi75E-G) using a commercial atomic force microscope (Park Systems NX10) combined with a lock-in amplifier (Stanford Research Systems SR830).

**Electrical and photoconductivity measurements.** Electrical transport measurements were performed along the in-plane direction of the LNO thin films. Au electrodes were patterned on the LNO thin film surface by thermal evaporation using a shadow mask with thickness of 100 nm, defining a channel 1000 μm long and 75 μm wide. For the light source, we used a UV laser of 325 nm (~3.8 eV) and intensity of 0.5 mW/cm$^2$.

## RESULTS & DISCUSSION

A systematic vacancy modulation was achieved in LiNbO$_3$ (LNO) thin films using pulsed laser deposition (PLD), while preserving the global hexagonal framework structure. Figure 1 shows the lattice and atomic structures of LNO (001) thin films deposited on SrTiO$_3$ (STO) (111) or Nb-doped STO substrates under different oxygen partial pressures ($P$(O$_2$)). Phase-pure LNO thin films were obtained between $P$(O$_2$) of 5 and 50 mTorr. All the thin films were highly oriented along the [00$l$] direction, as evidenced from x-ray diffraction (XRD) $\theta$-2$\theta$ scans (Figure 1(a)). To the best of our knowledge, LNO thin film growth on STO (111) substrates has not been reported previously. Ideally, an STO (111) substrate (in-plane lattice constant of 5.52 Å) imposes a rather large tensile strain (+6.7%) with a larger lattice constant compared to that of bulk LNO (5.15 Å), which prevents an epitaxial growth. Nevertheless, a highly oriented crystalline structure was realized for all of the LNO thin films used in this study, possibly owing to the same in-plane hexagonal symmetry and chemical strain, details of which will be discussed later. While preserving the overall lattice framework, subtle change in the structure could also be manifested. For example, the out-of-plane lattice spacing showed a systematic increase with increasing $P(O_2)$ (up to 30 mTorr), as shown in Figure 1(b). The large value of out-of-plane lattice constant of the thin film compared to that of the bulk (2.31 Å) can be attributed to the off-stoichiometric nature of the former, which is well-known for the PLD grown thin films.[27] In particular, Li vacancies have been widely exploited in LNO crystals by virtue of the light-weight nature of the element.[28-29]

Upon performing scanning transmission electron microscopy (STEM), we could confirm the hexagonal atomic structure and strained nature of the LNO thin films. Figures 1(c)–1(f) show annular bright field (ABF) STEM images of the LNO thin films deposited at $P(O_2)$ = 5 and 30 mTorr. The ABF STEM can directly visualize both light and heavy elements and show the respective atomic columns as dark features in the resulting image.[30] The images are shown for the $(1\bar{1}0)$ cross-sectional plane of the STO substrates. In the low-magnification images (Figures 1(c) and 1(d)), a coherently grown LNO thin film with hexagonal structure can be observed. Fast Fourier transform (FFT) patterns for the two films (insets) establish that the global lattice frameworks of the two films are well maintained (also see Figure S5). Note that the interface between the thin film and the substrate is sharply defined, but with an interface region of thickness ~1 nm. This interface layer can be attributed to a chemical intermixing layer (see Figure S1), and seems to facilitate the directional growth of the thin film despite the presence of large lattice mismatch. From the high-magnification images (Figures 1(e) and 1(f)), both the out-of-plane and in-plane lattice constants of the LNO thin films can be directly measured. The out-of-plane lattice constants obtained from STEM are consistent with those measured from the XRD. Surprisingly, however, the in-plane lattice constants of LNO thin films grown at 5 and 30 mTorr are identical (5.23 Å), despite the existence of the intermixing layer, suggesting that the thin films are at least partially strained. The 6-fold symmetry of the XRD $\varphi$-scan peaks around the LNO (104) reflections further supports this claim (see Figure S2). Therefore, the thin intermixing layer seems to transmit the strain and facilitate a coherent growth of the LNO thin film.[31] We note that the lattice expansion due to the non-stoichiometry and the presence of the intermixing layer let us discard the nominal lattice mismatch of 6.7% in our LNO thin films. The expanded lattice due to the chemical strain partially releases the tensile strain of the substrate, which facilitates the coherent growth of the film. On the other hand, STEM images clearly depict the displacement of Nb atoms along the out-of-plane direction, as expected in the normal polar crystal structure of bulk LNO (Figure 1(e)). The displacement of Nb together with that of Li could be regarded as the microscopic origin of ferroelectricity in LNO thin films. (Note that the actual displacement can be observed for Nb only, because Li has small projection column spacing (0.7 Å) which prevails beyond the resolving power of the current STEM technique.)

As the chemical stoichiometry plays a major role in determining the atomic and crystal structure, we proceeded to measure the relative atomic concentrations using x-ray photoemission spectroscopy (XPS). Figure 2 shows that Li concentration decreased, while Nb and O concentrations increased steadily with increase in $P(O_2)$ during film growth. We note that quantitative determination of stoichiometry is extremely difficult in thin films in general, but the qualitative trend is clearly observed for our LNO thin films. As $P(O_2)$ increased from 5 to 50 mTorr, the spectral weight of Li 1$s$ decreased progressively, as shown in Figure 2(a). A low Li concentration suggests that the light-weight Li atoms scatter easily in oxygen ambient during the deposition at high $P(O_2)$.[32] Nb and O concentrations were also estimated to increase with increasing $P(O_2)$ (see Figure S3). To investigate the spectra in further detail, we deconvoluted the peaks using a mixed Gaussian and Lorentzian functions, and present the relative atomic concentration of each element in Figure 2(b). Evaluation of the systematic trends in stoichiometry of LNO thin films has also been qualitatively verified using x-ray absorption spectroscopy (XAS). As shown in Figure S4, two main peaks near 2375 and 2379 eV correspond to the unoccupied Nb $4d$ ($Nb^{5+}$; $4d^0$) $t_{2g}$ and $e_g$ orbitals, respectively. As $P(O_2)$ decreases, low-energy peaks near 2374 and 2378 eV emerge, suggesting the prevalence of a lower valence state of Nb ion (i.e. $Nb^{4+}$; $4d^1$). The XAS result is consistent with the prospect of increasing Li and/or decreasing O concentration in the thin film.

The large modulation of Li concentration (from 19.8% to 11.2%) together with the relatively large Nb concentration (> 26%), allowed us to consider the inclusion of anti-site Nb defects, i.e., Nb ions replacing nominal Li ions at Li sites. Anti-site Nb defect has been frequently observed in LNO crystals, owing to the dissimilar bond strength and nearly equal ionic radii of the Li and Nb ions.[33] It has also been considered to describe the motion of ferroelectric domain walls.[34-35] Therefore, the stoichiometry of our LNO thin film let us consider a Li-deficient $LiNb_3O_8$ phase in accordance to the anti-site Nb defect, which was reported in few studies of PLD-grown thin films.[36-37] It is indeed well-known that the $LiNb_3O_8$ phase grows epitaxially in three dimensions within the LNO matrix with the orientation relationship of $(010)_{LiNb3O8} \parallel (2\overline{1}\overline{1}0)_{LNO}$ and $(\overline{1}01)_{LiNb3O8} \parallel (0001)_{LNO})$.[38] Therefore, we suggest that our LNO thin film grown at high $P(O_2)$ has both $LiNbO_3$ and $LiNb_3O_8$ phases induced by Li vacancies, but maintaining the hexagonal structural framework. These insights provide means to understand the modulation of polar behavior in LNO thin films.

The stoichiometry change leads to an unexpected, yet substantial rotation in the ferroelectric polarization direction in LNO thin films. The rotation of the polarization direction was revealed by measuring the structural symmetry breaking direction by means of second harmonic generation (SHG). Dipole-allowed SHG signal is generated by a polarized light applied on a crystal with broken inversion symmetry. We monitored the SHG response in three different configurations, i.e., $P_{in}P_{out}$, $S_{in}P_{out}$, and $S_{in}S_{out}$, where $P$ and $S$ with a subscript in (out) indicates a parallel and perpendicular polarization with respect to the plane of incidence for fundamental (second harmonic) light, respectively. Among these, $P_{in}P_{out}$ and $S_{in}P_{out}$ configurations are relevant for understanding the polarization rotation, and the sample azimuth dependent SHG intensity is shown in the inset of Figure 3(a). The experimental results from these two configurations exhibit a clear threefold modulation reflecting the 3$m$ point group symmetry of the LNO crystal. According to the symmetry, the SHG response in each configuration should follow the relation,

$$I_{PP}(2\omega) = \left(a_1 + a_2\left[-3\sin(\varphi)\cos^2(\varphi) + \sin^3(\varphi)\right]\right)^2, \quad (1)$$

$$I_{SP}(2\omega) = \left(b_1 + b_2\left[-3\sin(\varphi)\cos^2(\varphi) + \sin^3(\varphi)\right]\right)^2, \quad (2)$$

where $\varphi$ is the azimuth angle.[39] Note that $a_1$, $a_2$, $b_1$, and $b_2$ are given as functions of the susceptibility tensor components as, $a_1 = f(\chi_{zzz}, \chi_{zxx}, \chi_{xzx})$, $a_2 = f(\chi_{yyy})$ and $b_1 = f(\chi_{zxx})$, and $b_2 = f(\chi_{yyy})$, and hence $a_1$ and $b_1$ ($a_2$ and $b_2$) reflect the anisotropic symmetry breaking along the out-of-plane (in-plane) direction. Accordingly, the equations (1) and (2) imply that, when the crystal has predominantly out-of-plane symmetry breaking components, $a_1/a_2$ and $b_1/b_2$ are larger than 1, and the SHG response should show typical threefold symmetry. On the other hand, when the crystal has in-plane symmetry breaking components, $a_1/a_2$ and $b_1/b_2$ are smaller than 1, and the SHG response should show sixfold symmetry. The evolution of the SHG response from a threefold symmetry ($P(O_2)$ = 5 mTorr) to a sixfold symmetry ($P(O_2)$ = 50 mTorr) clearly suggests that the in-plane symmetry breaking components dominate the LNO thin film as more Li vacancies are introduced. The anisotropic symmetry breaking components obtained by fitting of SHG signal is shown in Figure 3(a). As expected, $a_1$ ($a_2$) decreases (increases) systematically, as $P(O_2)$ increases, indicating that a rotation of polarization from an out-of-plane direction to an in-plane direction is feasible. We note that the $a_1/a_2$ value for the bulk LNO single crystal ($a_1/a_2$ = 2.5) is considerably larger than that for the LNO thin film deposited at $P(O_2)$ = 5 mTorr ($a_1/a_2$ = 0.7).[39] This implies that even the thin film deposited at the lowest possible $P(O_2)$ possesses considerable amount of in-plane symmetry breaking components, an observation which is consistent with the non-zero Li vacancies observed in the film deposited at $P(O_2)$ = 5 mTorr, as shown in Figure 2(b). The LNO thin film deposited at $P(O_2)$ = 50 mTorr gives the ($a_1/a_2$) value of 0.04, indicating that the thin film has mostly in-plane symmetry breaking components.

The evidence for rotation of the inversion symmetry breaking direction could be corroborated using piezoresponse force microscopic (PFM) measurements in both out-of-plane and in-plane directions.[40] Indeed, the PFM images show highly anisotropic distribution of the polarizations. The PFM images of the as-prepared LNO thin film deposited at $P(O_2)$ = 5 mTorr in the upper insets of Figure 3(b) indicate the existence of both well-defined out-of-plane and in-plane polarizations. On the other hand, as shown in the lower insets, the as-prepared film deposited at $P(O_2)$ = 30 mTorr exhibits polarization aligned predominantly along the in-plane direction, and only a random distribution of the phases was observed along the out-of-plane direction. By counting the number of pixels on each PFM image, histograms can be drawn as shown in Figure 3(b). For the LNO thin film deposited at low $P(O_2)$, polarization direction can be considered as a combination of both out-of-plane and in-plane directions. On the other hand, for the film deposited at high $P(O_2)$, the polarization shows a strong peak along the in-plane direction only and is scattered along the out-of-plane direction, indicating noise-like phase pattern. The PFM result is remarkably consistent with the SHG signal, an observation which validates the substantial rotation of the ferroelectric polarization.

By examining different crystal structures within the LNO thin films, we could obtain an insight into the origin of the polarization rotation in the order-disorder-type ferroelectric. In particular, the large amount of Li deficiencies and relatively high Nb concentration in the LNO films deposited at high $P(O_2)$ prompted us to consider the Li-deficient $LiNb_3O_8$ phase with a monoclinic structure. Indeed, XRD pole figure measurements provide the evidence of monoclinic-like phase within the LNO thin films, as shown in Figure 4. When $2\theta = 32.396°$, the expected sixfold symmetry of $LiNbO_3$ (104) diffraction peaks is shown along with the threefold STO (110) diffraction peaks for the films deposited at low $P(O_2)$ (Figure 4(a)). On the other hand, at $2\theta = 30.200°$, a weak twin structure with a twofold symmetry could be observed, which is reminiscent of the $\bar{2}12$ Bragg reflection of the monoclinic $LiNb_3O_8$-like phase (Figure 4(b)). For the LNO films deposited at high $P(O_2)$, the peaks of the monoclinic-like phase grow in intensity (Figure 4(d)), and are much broader in the sense that a trace is even evident at $2\theta = 32.396°$ (Figure 4(c)). Moreover, instead of the clear sixfold symmetry diffraction peaks of the $LiNbO_3$ phase, twinned sixfold symmetry Bragg reflection peaks were seen at $2\theta = 30.200°$ (Figure 4(d)). The twinning of the crystal might be associated with partially relaxed LNO with double domain.[41] The twinned sixfold peaks have larger intensity compared to that of the twofold peaks which indicates that the main structure of the LNO films deposited at high $P(O_2)$ is still a hexagonal one. However, the hexagonal structure from the LNO thin film grown at high $P(O_2)$ is clearly different from the one from the film grown at low $P(O_2)$. The substantial difference in the Li and Nb concentration and the increased volume fraction of the monoclinic $LiNb_3O_8$-like phase distort the original hexagonal lattice, which seems to promote global in-plane ferroelectricity. The inclusion of local monoclinic phase could be further verified through STEM data. While the global crystal structure was identified to be the same as that of the FFT pattern (insets of Figures 1(c) and 1(d)), subtle monoclinic distortions could be manifested in LNO thin films with larger Li vacancies (see Figure S5).

The monoclinic $LiNb_3O_8$-like phase can indeed facilitate the unexpected in-plane polarization. In bulk, the monoclinic $LiNb_3O_8$ does not have any out-of-plane ionic displacement as observed for the hexagonal $LiNbO_3$ (See Figure S6). The absence of out-of-plane ionic displacement can also be explained in terms of anti-site Nb defect, which restores the inversion symmetry along the out-of-plane direction. Instead, an in-plane ionic displacement can be expected within the $P2_1/a$ structure. While the monoclinic $LiNb_3O_8$ structure belongs to the centrosymmetric point group, the spatial inversion symmetry breaking can occur owing to the off-centered ions along the in-plane direction. Especially, with the introduction of (partial) epitaxial tensile strain, lattice instability along the in-plane direction might be enhanced for the $LiNb_3O_8$-like phase. The spatial inversion symmetry breaking along the in-plane direction in the $LiNb_3O_8$-like phase seems to proliferate across the thin film, owing to the distorted hexagonal lattice. As the distortion of the hexagonal lattice is due to the inclusion of the monoclinic $LiNb_3O_8$ phase, it might naturally adopt the in-plane spatial inversion symmetry. After all, the structures are not that different. As a consequence, the breaking of mirror symmetry can be anticipated, evidenced by the observed twinned structure from the XRD pole figure measurements for the films grown at high $P(O_2)$.

The in-plane symmetry breaking in $LiNb_3O_8$-like phase would give rise to the in-plane ferroelectric polarization, consistent with our PFM and SHG results. Specifically, (i) the non-zero in-plane polarization for the thin films deposited at low $P(O_2)$, (ii) the increase of in-plane polarization with increasing $P(O_2)$, and (iii) the decrease of out-of-plane polarization with increasing $P(O_2)$ can all be understood in terms of enlarged monoclinic-like phase within the LNO thin film due to absence of Li ions and following structural distortion of the original hexagonal phase. This leads to rotation of the direction of the spontaneous polarization in the order-disorder type ferroelectric LNO.

Interestingly, the polarization rotation observed in the LNO thin films is fundamentally different from that observed in conventional displacive-type ferroelectrics. In displacive-type ferroelectrics, the polarization rotation occurs discontinuously at an MPB with high electromechanical response, leading to an enhanced piezoelectric and dielectric response. On the other hand, in order-disorder-type ferroelectrics, or at least in our case with LNO thin film, the polarization rotation occurs continuously. The absence of a discontinuous rotation and MPB suggests the absence of the maximum in the dielectric response in the order-disorder type ferroelectrics, consistent with our experimental observation. Indeed, we observe qualitatively similar degree of symmetry breaking (Figure 3(a)) and piezoelectric response (Figure 3(b) and Figure S7) by changing the polarization direction.

The rotation of the spontaneous polarization direction may have wide implication in various electronic phenomena including the ferroelectric photovoltaic effect.[18] In particular, an enhanced photoconduction with hysteretic behavior was observed during in-plane photoconductivity measurements on the LNO thin film deposited at high $P(O_2)$. Figure 5 shows the current-voltage (I-V) characteristics of the LNO thin films with the in-plane geometry. The inset in Figure 5(a) is a schematic of measurement setup. The LNO thin film deposited at low $P(O_2)$ exhibited a limited photo-response enhancement under illumination with UV light, as shown in Figure 5(a). The ratio between the light and dark current was less than two. Thus, the photoconductivity of LNO was moderate when the in-plane ferroelectric polarization was relatively small. On the other hand, for the film deposited at high $P(O_2)$, a significantly enhanced photoconductivity was observed with the ratio $I_{light}/I_{dark} = 10$, as shown in Figure 5(b). In addition, a hysteretic I-V curve was obtained, manifesting the effect of the in-plane ferroelectric polarization. We note that the dip in the I-V curve corresponds to a field of ~1 kV/cm, which is about an order of magnitude smaller than the coercive field reported for polycrystalline LNO films measured along the out-of-plane direction.[42] We believe that such a discrepancy might arise from the superior crystallinity of our thin film samples and also the in-plane geometry. We further note that the overall low current level, which is advantageous for ferroelectric applications, might be associated with low oxygen vacancy concentration in the case of films with high Li vacancy concentration.

**SUMMARY**

In summary, the feasibility of rotating the direction of the spontaneous polarization in the order-disorder type ferroelectric $LiNbO_3$ thin films has been demonstrated by varying the elemental stoichiometry and resultant local lattice structure in the thin film. By increasing the oxygen partial pressure during the pulsed laser deposition procedure, we could systematically introduce Li vacancies, which lead to the partial formation of $LiNb_3O_8$-like phase. The nominal ferroelectric polarization along the out-of-plane direction, expected for the hexagonal $LiNbO_3$, tilted systematically toward the in-plane direction with accompanying increase in the volume fraction of local monoclinic $LiNb_3O_8$-like phase in the thin film. Both the second harmonic generation and piezoelectric force microscopy data consistently supported the rotation of the spontaneous polarization direction, based on the difference in the inversion symmetry breaking direction and anisotropic distribution of the piezoelectric coefficients, respectively. Our findings may prove to be beneficial in comprehending the ferroelectric phenomena in $LiNbO_3$, especially for engineering the large ferroelectric polarization by means of charge-lattice coupling in this order-disorder-type ferroelectric material.

FIGURES.

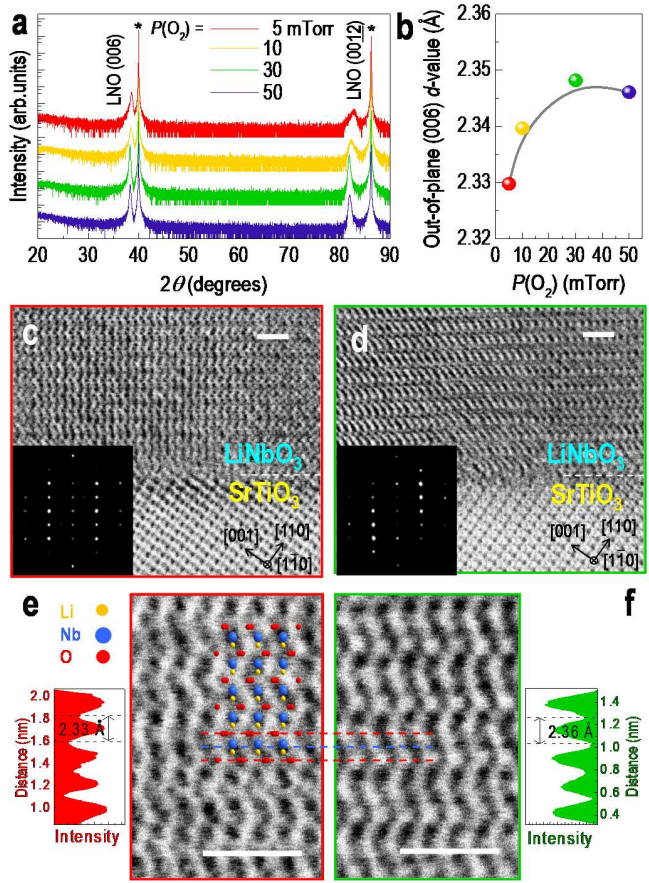

**Figure 1.** Lattice and atomic structure of LiNbO$_3$ thin films deposited under various oxygen partial pressures. (a) X-ray diffraction $\theta$-$2\theta$ scans of phase-pure LiNbO$_3$ thin films deposited on SrTiO$_3$ (111) substrate. (b) Change in the *c*-axis lattice constant as a function of oxygen partial pressure. Low-resolution bright-field scanning transmission electron microscopic images of the LiNbO$_3$ thin film deposited at oxygen partial pressure of (c) 5 and (d) 30 mTorr. The insets show fast Fourier-transform diffraction patterns of the corresponding images. High-magnification images of the films deposited at (e) 5 and (f) 30 mTorr, separately obtained for each sample. The overall LiNbO$_3$ structure is preserved regardless of different oxygen partial pressures used. The scale bars represent 1 nm.

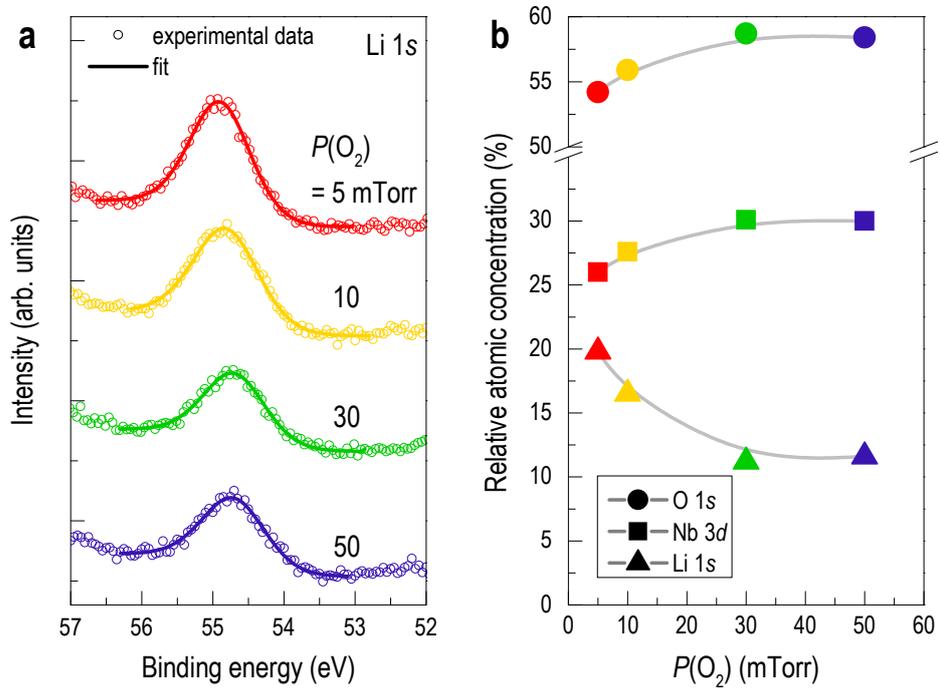

**Figure 2.** Relative atomic concentration of the constituent elements in LiNbO$_3$ thin films. (a) X-ray photoelectron spectroscopy near the Li 1*s* state. With decreasing oxygen partial pressure during the growth, reduction of binding energy is observed, indicating the introduction of Li vacancies. (b) Relative atomic concentrations (%) as a function of oxygen partial pressure during thin film deposition revealing a systematic change in the film stoichiometry.

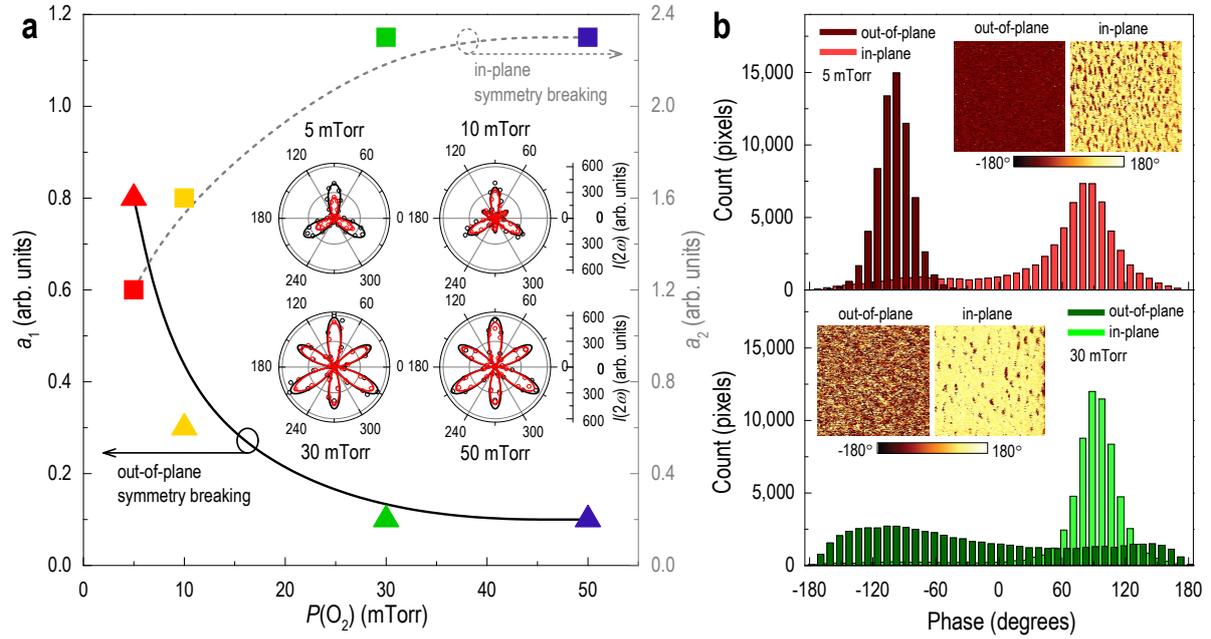

**Figure 3.** Anisotropic symmetry breaking and polarization rotation. (a) Anisotropic symmetry breaking components, $a_1$ and $a_2$, calculated from curve-fitted second harmonic generation (SHG) data for the $P_{in}P_{out}$ (black) and $S_{in}P_{out}$ (red) polarization configurations, respectively. $a_1$ and $a_2$ represent symmetry breaking components along the out-of-plane and in-plane directions, respectively. The inset shows the experimental (symbols) and fit (lines) of the SHG in LiNbO$_3$ thin films indicating strong anisotropy in structure symmetry breaking. (b) Distribution of the polarization phases are shown for the out-of-plane and in-plane PFM images of LiNbO$_3$ thin films deposited at 5 mTorr (upper panel) and 30 mTorr (lower panel) of oxygen partial pressures, respectively. Insets show anisotropic PFM images. The size of the image is 0.5 × 0.5 $\mu$m$^2$.

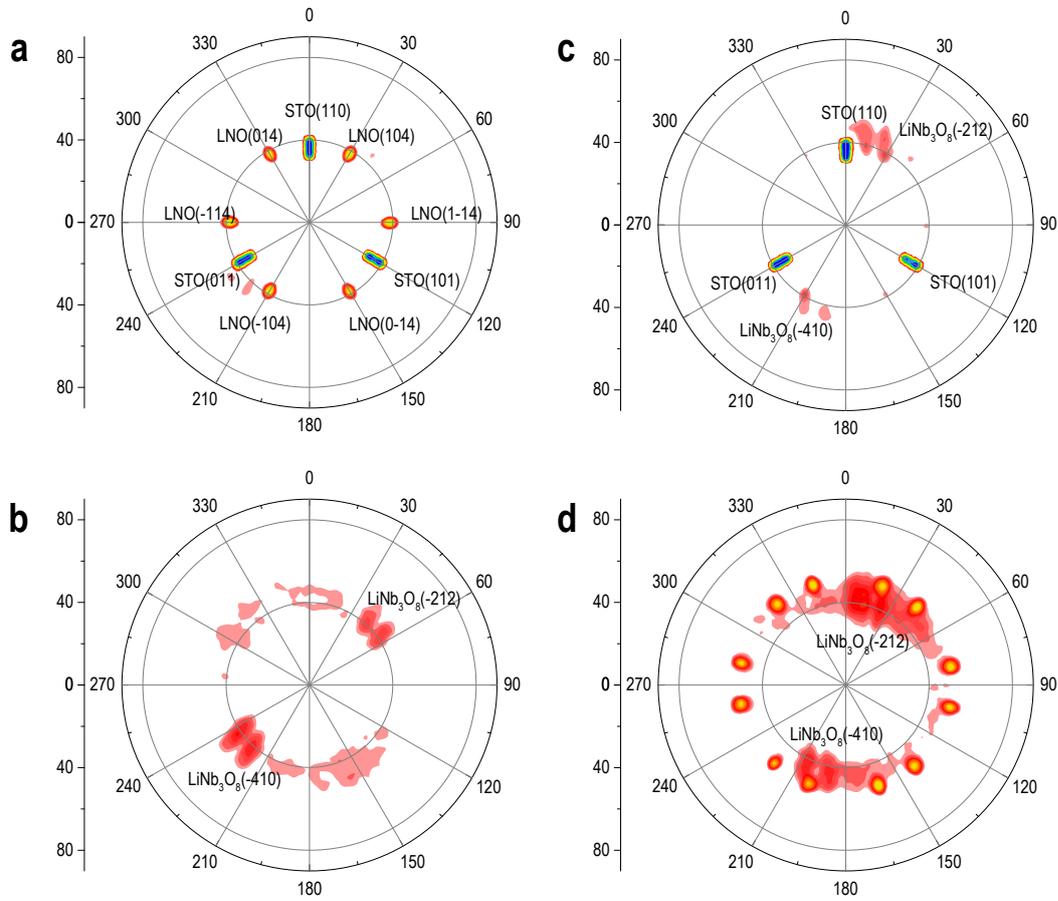

**Figure 4.** XRD pole figure analysis. Pole figure measurements on LiNbO$_3$ thin film samples deposited at $P(O_2)$ = (a,b) 5 and (c,d) 30 mTorr. Two different $2\theta$ configurations (a,c) 32.396° and (b,d) 30.200° were used. Hexagonal LiNbO$_3$ thin film phase with sixfold symmetry, monoclinic LiNb$_3$O$_8$-like phase with twofold symmetry, and cubic SrTiO$_3$ substrate phase with threefold symmetry can be identified depending on the $P(O_2)$ and $2\theta$ values.

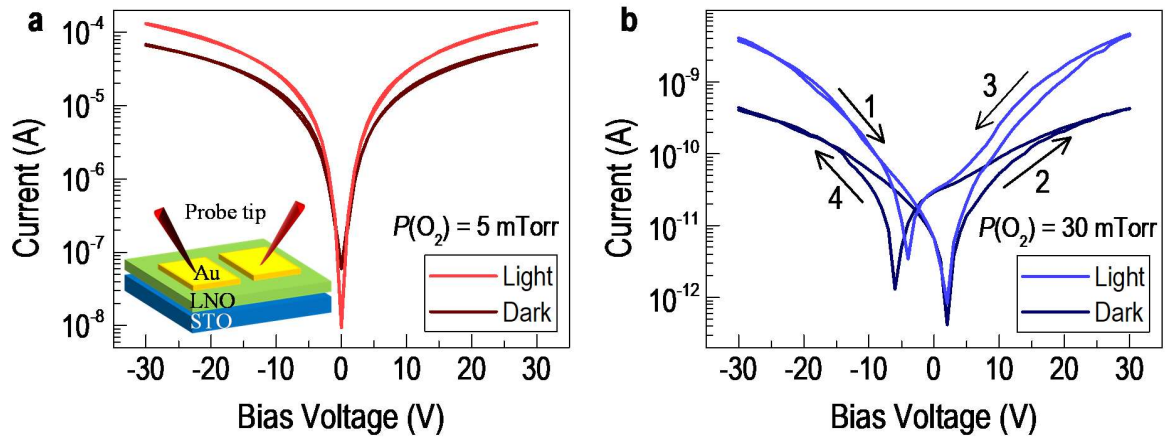

**Figure 5.** Anisotropic photoconductivity in LiNbO$_3$ thin films. In-plane photoconductivity of the LiNbO$_3$ thin films deposited at (a) 5 mTorr and (b) 50 mTorr. The large in-plane spontaneous polarization in LiNbO$_3$ thin film deposited at high $P$(O$_2$) results in hysteresis and enhanced photoconductivity.

ASSOCIATED CONTENT

**Supporting Information**.

Chemical intermixing layer at the interface between LiNbO$_3$ thin film and SrTiO$_3$ substrate; Modulation of atomic concentration in LiNbO$_3$ thin films; Evidence of stoichiometry change observed from x-ray absorption spectroscopy; Evidence of monoclinic distortion with introduction of Li vacancies in LiNbO$_3$ thin films; Understanding polarization rotation with inclusion of monoclinic LiNb$_3$O$_8$ phase. This material is available free of charge.

AUTHOR INFORMATION

**Corresponding Author**

*E-mail: choiws@skku.edu.*E-mail: choiws@skku.edu.

**Author Contributions**

The manuscript was written through contributions of all authors. All authors have given approval to the final version of the manuscript. ‡These authors contributed equally.

ACKNOWLEDGMENT

We thank J. Lee for insightful discussion. This work was supported by Basic Science Research Programs through the National Research Foundation of Korea (NRF) (NRF-2017R1A2B4011083, NRF-2016R1A6A3A11934867 (S.A L.), NRF-2015R1A5A1009962 (C.R. & J.L.), NRF-2014R1A4A1008474 (S.K., D.S., and Y.K.), NRF-2015R1D1A1A01058672 (J.K. & S.P.), NRF-2015R1C1A1A02037514 (D.-Y.C.), NRF-2016R1D1A1A02937051 (J.Y.J.), and NRF-2015M3D1A1070672 (Y.M.K.). Y.M.K. was also supported by the Institute for Basic Science (IBS-R011-D1) ).

# Ferroelectric Polarization Rotation in Order-Disorder-Type LiNbO$_3$ Thin Films

*Tae Sup Yoo[1,‡], Sang A Lee[1,‡], Changjae Roh[2], Seunghun Kang[3], Daehee Seol[3], Xinwei Guan[4], Jong-Seong Bae[5], Jiwoong Kim[6], Young-Min Kim[7,8], Hu Young Jeong[9], Seunggyo Jeong[1], Ahmed Yousef Mohamed[10], Deok-Yong Cho[10], Ji Young Jo[11], Sungkyun Park[6], Tom Wu[4,12], Yunseok Kim[3], Jongseok Lee[2], and Woo Seok Choi[1,*]*

[1]Department of Physics, Sungkyunkwan University, Suwon 16419, Korea

[2]Department of Physics and Photon Science, Gwangju Institute of Science and Technology (GIST), Gwangju 61005, Korea

[3]School of Advanced Materials Science and Engineering, Sungkyunkwan University, Suwon 16419, Korea

[4]Materials Science and Engineering, King Abdullah University of Science and Technology, Thuwal 23955-6900, Saudi Arabia

[5]Busan Center, Korea Basic Science Institute, Busan 46742, Korea

[6]Department of Physics, Pusan National University, Busan 46241, Korea

[7]Department of Energy Science, Sungkyunkwan University, Suwon 16419, Korea

[8]Center for Integrated Nanostructure Physics, Institute for Basic Science (IBS), Suwon 16419, Korea

[9]UNIST Central Research Facilities, Ulsan National Institute of Science and Technology, Ulsan

44919, Korea

[10]IPIT & Department of Physics, Chonbuk National University, Jeonju 54896, Korea

[11]School of Materials Science and Engineering, Gwangju Institute of Science and Technology (GIST), Gwangju 61005, Korea

[12]School of Materials Science and Engineering, University of New South Wales, Sydney, NSW 2052, Australia

[*]e-mail : choiws@skku.edu

**Chemical intermixing layer at the interface between LiNbO₃ thin film and SrTiO₃ substrate**

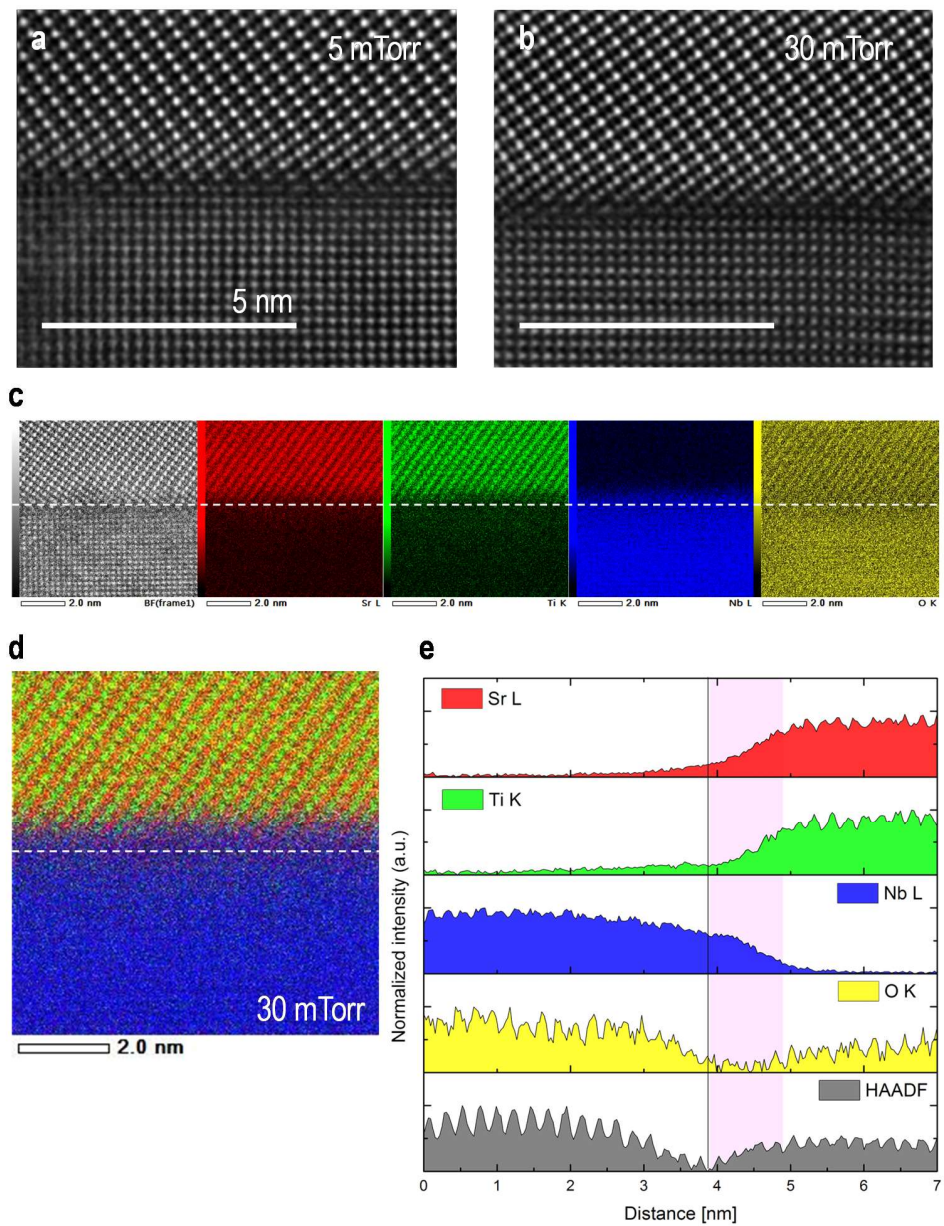

**Figure S1.** Scanning transmission electron microscopy (STEM) with energy dispersive X-ray analysis (EDX). High-resolution STEM images near the interface for the LNO thin films grown at (a) 5 and (b) 30 mTorr of $P(O_2)$. (c) STEM images and elemental identification by EDX (total image (gray), Sr (red), Ti (green), Nb (blue), O (yellow)). (d) Total image created by gathering the data from (c). Horizontal dashed line represents the interface. (e) Normalized elemental intensity across the interface.

The nature of interfaces between the LiNbO$_3$ thin films and SrTiO$_3$ substrates were assessed by scanning transmission electron microscopy (STEM). In order to identify the interface region, we used scanning transmission electron microscopy in conjunction with energy dispersive x-ray analysis (STEM-EDX), which revealed a chemical intermixing layer as shown in Fig. S1. Figure S1a shows the STEM image in which each atom is marked with a different color. At the same time as the data were collected, we obtained Fig. S1b that revealed the presence of Nb atoms near the surface of the STO substrate. The intermixing layer was estimated to be ~1 nm in thickness, as shown in Fig. S1c. We further note that there is a possibility that this intermixing layer might have formed during the TEM sample preparation (ion milling), and not during the thin film growth. Selective etching can occur in heterostructures composed of materials having large differences in mechanical properties, which could lead to interface amorphization as seen in GaN thin films deposited on sapphire substrates.[1,2]

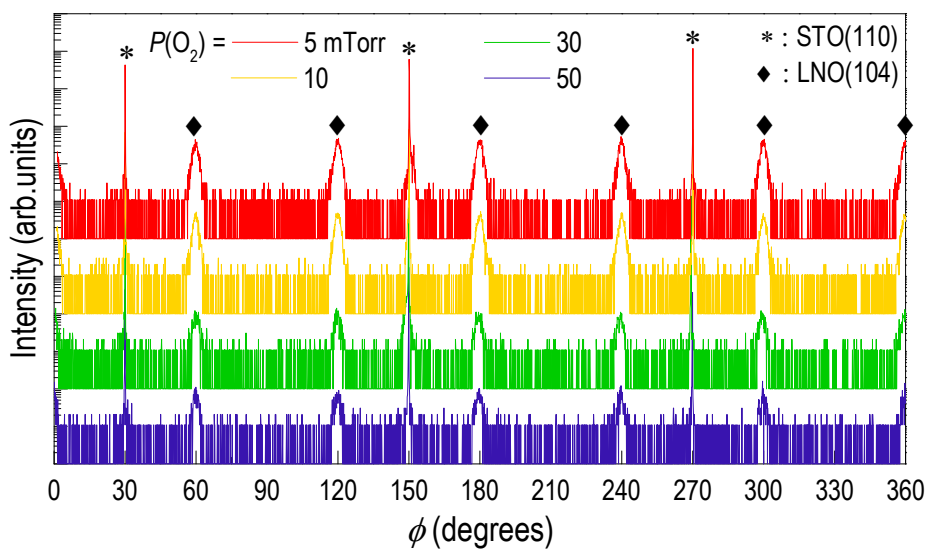

**Figure S2.** X-ray diffraction (XRD) $\varphi$-scan around LNO (104) Bragg reflection. All of the LNO thin films show hexagonal symmetry.

**Modulation of atomic concentration in LiNbO₃ thin films**

Relative atomic concentration in the LiNbO₃ (LNO) thin film was obtained from the spectral weight of the corresponding elemental binding energy peak in the x-ray photoelectron spectra (XPS). As shown in Fig. S3(a), a modulation of atomic content was observed for different oxygen partial pressures during the growth, $P(O_2)$. The spectral weight of each elemental species is shown in Figs. S3(b), (c), and (d), for Li 1$s$, Nb 3$d$, and O 1$s$, respectively.

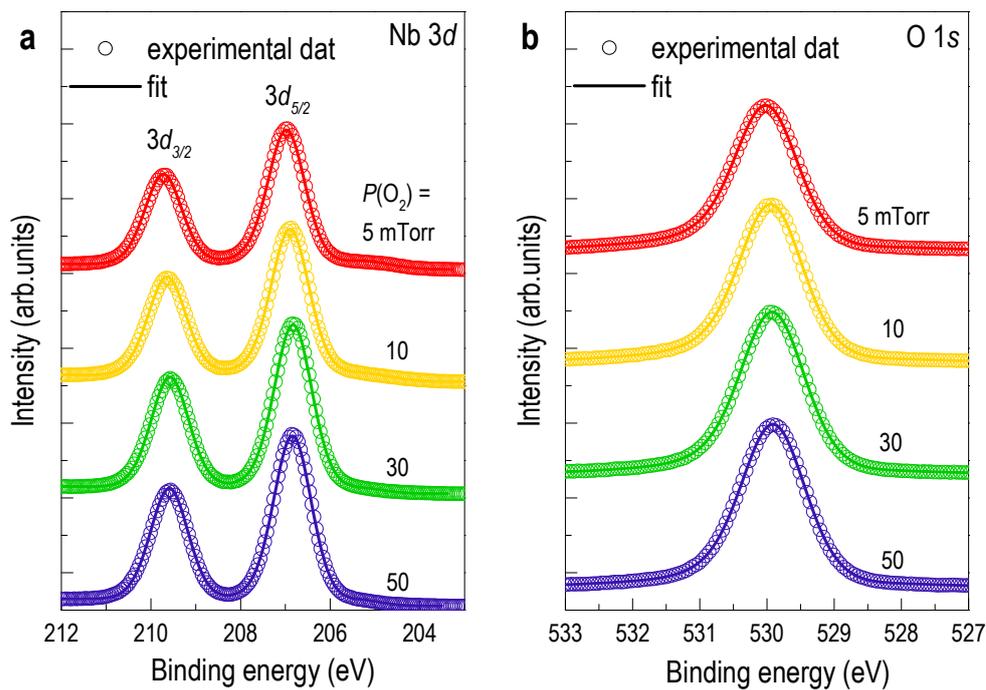

**Figure S3.** X-ray photoelectron measurements. XPS spectra of LNO thin films deposited at different $P(O_2)$. (a) Nb 3$d$ and (b) O 1$s$.

**Evidence of stoichiometry change observed from x-ray absorption spectroscopy**

X-ray absorption spectroscopy (XAS) yields electronic band structure of valence band, and provides additional information regarding the stoichiometry of LNO thin films. Figure S4 shows the XAS spectra near the Nb $L_3$ absorption edge. For stoichiometric LNO, nominal $Nb^{5+}$ valence state is expected. However, as Li and oxygen vacancy concentrations increase, $Nb^{4+}$ states are also introduced which is discernible in the XAS spectra of the LNO thin films deposited at low $P(O_2)$. This qualitative trend due to the $P(O_2)$ dependence corroborates the relative stoichiometric data obtained from the XPS measurements.

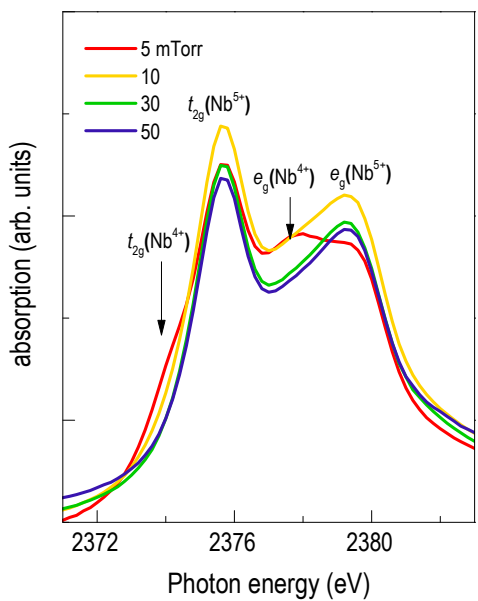

**Figure S4.** X-ray absorption spectra of LNO thin films. With decreasing $P(O_2)$, the absorption edge due to $Nb^{4+}$ becomes more visible in addition to that due to $Nb^{5+}$, indicating increasing concentrations of Li and/or oxygen vacancies.

**Evidence of monoclinic distortion with introduction of Li vacancies in LNO thin films**

Detailed analyses of the fast Fourier transform (FFT) data obtained from STEM images provide additional evidence for the inclusion of the monoclinic LiNb$_3$O$_8$-like phase. Figures S5(a) and S5(b) show the FFT patterns obtained from the STEM images (given in Figs. 1(c) and 1(d) earlier) for $P(O_2)$ = 5 mTorr) and 30 mTorr, respectively. When merged together (see Fig. S5(c)), the patterns coincide, indicating that the global crystal structure is maintained. However, upon magnification (Fig. S5(d)), a systematic shift of the diffraction spots is for the thin film deposited at $P(O_2)$ = 30 mTorr. This in-plane shift suggests that a slight or local monoclinic distortion exists in the LNO thin film with Li vacancies, consistent with the XRD results given in Fig. 4.

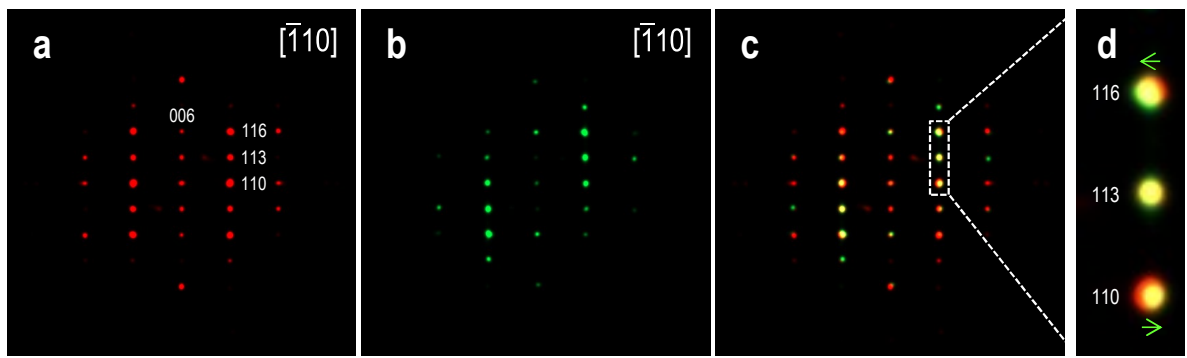

**Figure S5.** Subtle monoclinic distortion observed in LNO thin films deposited at high $P(O_2)$. Fast Fourier transform (FFT) data obtained from STEM images of LNO thin films deposited at (a) $P(O_2)$ = 5 mTorr and (b) 30 mTorr, respectively. (The corresponding STEM images are illustrated in Figs. 1(c) and 1(d)). (c) Merged FFT patterns show that the global crystalline structure is not disturbed. (d) Upon magnification, a systematic shift of the diffraction spots can be observed. This indicates slight monoclinic distortion in the LNO thin film deposited at high $P(O_2)$.

**Understanding polarization rotation with inclusion of monoclinic LiNb$_3$O$_8$ phase**

In order to explain polarization rotation with the inclusion of Li-vacancies, we invoked Li-deficient LiNb$_3$O$_8$ phase of monoclinic structure. The first evidence of the LiNb$_3$O$_8$ phase came from x-ray diffraction pole figure measurement. The second evidence came from the STEM analyses. Bulk LiNbO$_3$ has hexagonal structure with spontaneous polarization along the *c*-axis direction at room temperature as shown in Figs. S6(a) and S6(b). The structure has sixfold symmetry along in-plane direction. On the other hand, monoclinic structure has no ionic displacement along the out-of-plane direction, but exhibits local symmetry breaking with off-centered atoms along the in-plane direction (Figs. S6(c) and S6(d)). Although the off-centered atoms are not fully aligned, it is possible for polarization to occur with inclusion of defects or under application of external electric field. Since our LNO thin films are additionally subject to tensile strain, further space for atomic displacements could be provided.

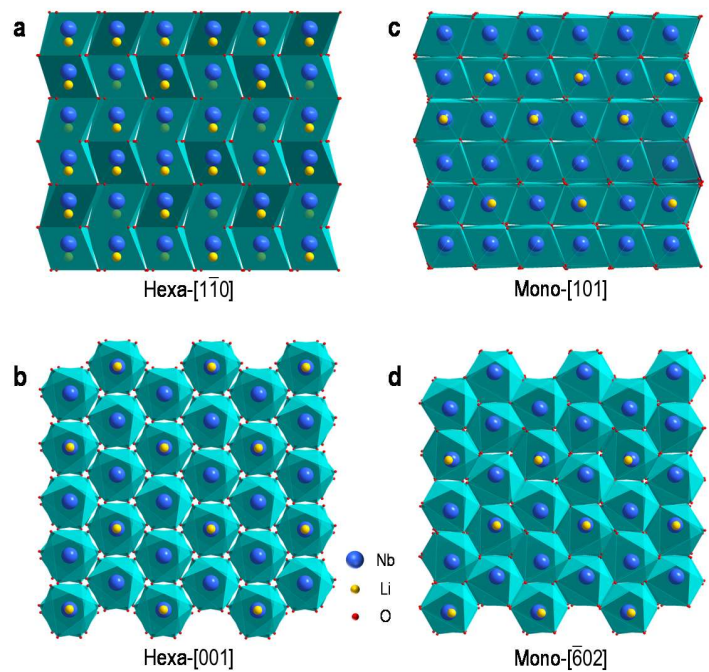

**Figure S6.** Comparison of LiNbO$_3$ and LiNb$_3$O$_8$ structure. Hexagonal LiNbO$_3$ ((a) and (b)) shows ionic displacement and asymmetry along the out-of-plane direction ((a)). In contrast, monoclinic LiNb$_3$O$_8$ ((c) and (d)) shows highly symmetric structure along the out-of-plane direction ((c)) and symmetry breaking with off-centered ions along the in-plane direction ((d)).

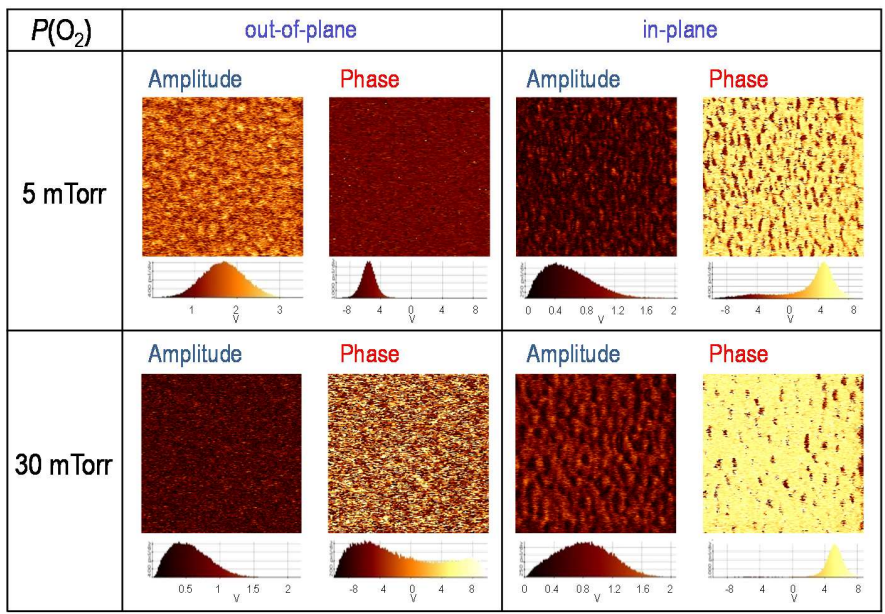

**Figure S7.** PFM images and distribution of the polarization amplitude and phase for the out-of-plane and in-plane of LiNbO$_3$ thin films deposited at 5 and 30 mTorr of $P(O_2)$. The sizes of the images are $0.5 \times 0.5$ μm$^2$.